\documentclass[aps,prd,nofootinbib,showpacs,twoside,notitlepage]{revtex4-1}

\usepackage{amsmath}
\usepackage{amssymb}
\usepackage{amsfonts}

\begin{document}


\title{Hamiltonian formalism and constraint analysis of three-form matter models coupled with general relativity}

\author{David Brizuela}
\email{david.brizuela@ehu.eus}
\affiliation{Fisika Teorikoa eta Zientziaren Historia Saila, UPV/EHU,
  644 P.K., 48080 Bilbao, Spain} 
  
\author{I\~naki Garay}
\email{inaki.garay@ehu.eus}
\affiliation{Fisika Teorikoa eta Zientziaren Historia Saila, UPV/EHU,
  644 P.K., 48080 Bilbao, Spain}


\begin{abstract}
A Hamiltonian analysis of models given by a three-form field with a generic potential coupled to general relativity
in four dimensions is performed. This kind of fields are naturally present in string theory and cosmological
scenarios. In particular, the action that will be considered has been extensively used during the last years
to propose inflationary and dark energy models. Nevertheless, in order to keep the discussion as generic as
possible, neither symmetries nor a specific form of the potential for the three-form field will be imposed.
Interesting and relevant results about the number of dynamical degrees of freedom of these models are obtained.
In addition, the analogy with a weakly-coupled scalar field is discussed. Finally, as a particular example
of this generic framework, the homogeneous and isotropic cosmological case will be presented.
\end{abstract}


\maketitle

\section{Introduction}

The use of gravitational models described by an action involving nonscalar fields (p-forms in general) is extended in the
literature both of fundamental theories of quantum gravity and of classical modifications to general relativity.
More concretely, in string theory, forms are mathematical objects that, in principle, may appear naturally
in effective low-energy actions. In this sense, the addition of forms to the action can be properly motivated.
In fact, during the last years the study of gauge three-forms in effective theories in four dimensions has
gained interest due to the physical effects they may have, such as the generation and neutralization of the
cosmological constant \cite{Duff:1980qv,Koivisto:2009sd,Farakos,Bandos:2018gjp}.  

In cosmology, the consideration of p-form fields has recently got an increasing relevance, specifically
in the context of inflationary theories and dark-energy models
\cite{Germani:2009iq, Germani:2009gg, Koivisto:2009sd, Kobayashi:2009hj, Koivisto:2012xm, Koivisto:2009fb,
Koivisto:2009ew, Yao:2017enb, Gruzinov:2004rq, DeFelice:2012jt, DeFelice:2012wy, Mulryne:2012ax, Kumar:2014oka, Kumar:2016tdn, Morais:2016bev}.
Traditionally, the way to produce inflation (the dramatic accelerated expansion of the universe at an early stage) was using one
or more scalar fields (inflatons). Nevertheless, it is unclear what these inflatons really are. In fact, since fundamental scalar
fields have not been observed yet, we could consider that inflation may have been produced by nonscalar fields with vectorial
or tensorial nature.

Since the pioneering work by Ford \cite{Ford89}, where a model of inflation driven by a vector field was proposed,
the possibility of including other kind of fields has been extensively explored. Although the vector-field models may provide
solutions to the coincidence problem or to the possible electromagnetic origin of the cosmological constant, most of them
encounter instabilities \cite{vectorinstabilities}.
Concerning higher-spin fields, at first they were
overlooked because they generically induce an anisotropy in the cosmological scenario, but due to unexpected temperature
anisotropies observed in the CMB they have gained attention again. In fact,
simple and viable inflationary models driven by a form field have been found \cite{Koivisto:2009sd}.

In particular, three-form fields have been used within the cosmological scenario not only to describe the early
accelerated expansion of the universe, but also its late-time speed up. For instance,
in certain cases it is
possible to generate k-essence models using three-forms \cite{gruzinov}. Furthermore,
it has also been shown that three-form models can induce abrupt events in the future
for some wide family of potentials, like the so-called \emph{little sibling of the big rip}  
\cite{Morais:2016bev,Bouhmadi-Lopez:2016dzw,BouBriGar,Morais:2016bev,Albarran:2015cda,BouhmadiLopez:2009pu,Bouhmadi-Lopez:2014cca}.

Given the interest on this kind of models, both for cosmology as well as for high energy physics,
in this paper their canonical formalism will be considered. More concretely, the Hamiltonian
formalism of a three-form field coupled to general relativity in four dimensions will be studied.
The discussion will be kept completely general, without any imposition of symmetries nor specific form of the potential.
The homogeneous and isotropic cosmological scenario will arise as a particular application of the theory.

The canonical study of the dynamics of these systems provides an interesting and complementary
view to the more usual Lagrangian framework. This Hamiltonian formalism, with the algebraic structure of the constraints,
displays the different dynamical degrees of freedom, clearly distinguishing the ones that are physical from those that are pure gauge.
In addition, this analysis eases the comparison with gravitational
systems given by other matter contents. More concretely, the analogy
of three-form fields with a weakly-coupled scalar field will be analyzed in detail below.
Furthermore, the Hamiltonian formalism presented here may be useful in order to analyze the initial-value
problem and to find numerical solutions for this model.
Finally, although much beyond the goal of this article, this treatment
paves the way towards a possible canonical quantization of the system.

The structure of the paper is the following. In Sec. \ref{secLegtransf} the action of the three-form model is introduced and, after a Legendre transformation,
the Hamiltonian and the constraints of the system are obtained. 
In Sec. \ref{cantransf}, firstly the usual dual decomposition of the three-form field is considered,
and then a suitable canonical transformation that makes some of the constraints first-class is performed.
At the end of this section, the constraint algebra is computed and
the analogy of this model with the one given by a weakly-coupled scalar field is discussed.
In Sec. \ref{cases} the obtained results are applied to the study of three different particular examples:
the singular case of a constant potential for the three-form field, the quadratic potential
(that in some cases is analogous to the scalar field model), and the homogeneous cosmological scenario.
Finally, Sec. \ref{conclusions} presents the main conclusions and results.

\section{Legendre transformation}\label{secLegtransf}

Let us consider a three-form matter field $A_{\mu\nu\rho}$ coupled with gravity, ruled by the following action \cite{Koivisto:2009sd,Germani:2009gg,Koivisto:2009fb}:
\begin{equation}
 S=\int d^4x \sqrt{-g}{\cal L}=\int dt L=\int dt\int d^3x \sqrt{-g} \left[R-\frac{1}{48}F^{\mu\nu\rho\sigma}F_{\mu\nu\rho\sigma}-V(A^{\mu\nu\rho}A_{\mu\nu\rho})\right],
\end{equation}
where units such that $16\pi G/c^4=1$ have been chosen, and
\begin{equation}
 F_{\mu\nu\rho\sigma}:=4\nabla_{[\mu}A_{\nu\rho\sigma]}=\nabla_{\mu}A_{\nu\rho\sigma}-\nabla_{\sigma}A_{\mu\nu\rho}+\nabla_{\rho}A_{\sigma\mu\nu}-\nabla_{\nu}A_{\rho\sigma\mu}.
\end{equation}
Greek indices are spacetime indices, and $R$ is the Ricci scalar. The conjugate momentum of $A_{\mu\nu\rho}$ is defined by computing variations of the Lagrangian with respect to time derivatives of
the three-form,
\begin{equation}
 \Pi^{\mu\nu\rho}=\frac{\delta L}{\delta\partial_0A_{\mu\nu\rho}}=-\frac{\sqrt{-g}}{6}F^{0\mu\nu\rho}.
\end{equation}
Due to the antisymmetry of the field-strength tensor $F^{\mu\nu\rho\alpha}$,
this definition implies the following three primary constraints of this model:
\begin{equation}\label{primary1}
 \Pi^{0ij}=0,
\end{equation}
where Latin letters stand for spatial indices. Therefore, the components $A_{0ij}$ of the three-form field are nondynamical.
Furthermore, concerning the geometric sector, one also obtains the usual four primary constraints of vacuum general relativity;
namely, the vanishing of the conjugate momenta of the lapse $N$ and shift $N^i$,
\begin{equation}\label{primary2}
\widetilde{P}^N=0,\qquad \widetilde{P}^{\vec{N}}_i=0.
\end{equation}
These and other objects will be defined with a tilde because, afterwards, a canonical transformation
to nontilded variables will be introduced.

By performing a Legendre transformation, one
gets the following Hamiltonian density:
\begin{equation}\label{hamiltonian}
{\cal H}=N \left[C +\frac{3}{\sqrt{h}}\Pi_{ijk}\Pi^{ijk}+\sqrt{h}V(A^2)\right]+N^i C_i+3 \Pi^{ijk}D_k A_{0ij},
\end{equation}
where $h$ is the determinant of the spatial metric $h_{ij}$, $D_i$ the covariant derivative compatible with it,
and we have defined the shorthand $A^2=A^{\mu\nu\rho}A_{\mu\nu\rho}$. In addition,
$C$ and $C_i$ are the usual Hamiltonian and diffeomorphism constraints of vacuum general relativity,
\begin{eqnarray}
C&=&-\sqrt{h}R^{(3)}+\frac{1}{\sqrt{h}}\pi^{ij}\pi_{ij}-\frac{\pi^2}{2\sqrt{h}}\,,\\
C_i&=&-2D_j\pi^j_i\,,
\end{eqnarray}
where $\pi^{ij}$ is the canonical conjugate momentum of the spatial metric $h_{ij}$,
given by $\pi^{ij}=\sqrt{h}(K^{ij}-Kh^{ij})$, with $K_{ij}$ the extrinsic curvature associated with $h_{ij}$,
and $K:=K^{ij}h_{ij}$. In addition, the shorthand $\pi:= \pi^{ij}h_{ij}$ stands for the trace of the momentum,
and $R^{(3)}$ is the Ricci scalar associated with $h_{ij}$.

At this point one can construct the primary Hamiltonian, which will be used to compute the evolution of different
objects from here on, by adding to $\eqref{hamiltonian}$ the primary constraints with arbitrary coefficients,
\begin{equation}
{\cal H}_p=N \left[C +\frac{3}{\sqrt{h}}\Pi_{ijk}\Pi^{ijk}+\sqrt{h}V(A^2)\right]+N^i C_i+3 \Pi^{ijk}D_k A_{0ij}
+\alpha \widetilde{P}^N+\alpha^i \widetilde{P}^{\vec{N}}_i+\alpha_{ij}\Pi^{0ij},
\label{Hamp}
\end{equation}
where $\alpha_{ij}$ is an antisymmetric matrix and thus contains three independent functions.

By evolution of the primary constraints, one obtains seven additional constraint equations. On the one hand,
three corresponding to the matter sector,
\begin{equation}\label{gauss}
D_k\Pi^{ijk}-\frac{N}{3}\sqrt{h}\frac{\partial V}{\partial A_{0ij}}=0,
\end{equation}
and, on the other hand, the Hamiltonian and diffeomorphism constraints,
\begin{eqnarray}\label{hamiltonian_constraint}
C+\frac{3}{\sqrt{h}}\Pi_{ijk}\Pi^{ijk}+\sqrt{h}V +N\sqrt{h}\frac{\partial V}{\partial N}=0,\\\label{diff_constraint}
C_i+N \sqrt{h}\frac{\partial{V}}{\partial N^i}=0.
\end{eqnarray}
Note that, as opposed to other matter models usually considered in the literature,
the coupling of the three-form field with gravity produces a nontrivial change of these
constraints, in particular making them dependent on the lapse and shift functions.

In a next step, one should compute the evolution of constraints \eqref{gauss}--\eqref{diff_constraint}
in order to check whether further constraints must be considered to get a closed set of constraints
under evolution. Nonetheless, it can be shown
that the evolution of these secondary constraints only impose conditions on the coefficients
$\alpha$, $\alpha_i$ and $\alpha_{ij}$, but do not imply any tertiary constraint. At this point,
we will refrain from giving these conditions explicitly, but will do so for the adapted variables
that will be introduced in the next section. As summary of this section, we have obtained the Hamiltonian
of the considered model \eqref{Hamp} and a complete set of constraints, \eqref{primary1},\eqref{primary2} and \eqref{gauss}--\eqref{diff_constraint}, which are closed under evolution. 

\section{Canonical transformation}\label{cantransf}

This section is composed by four subsections. In the first one, the usual decomposition of the three-form field in
terms of its dual variables will be performed. This is a very natural set of variables, since the four independent
components of the three-form field $A_{\mu\nu\rho}$ will be encoded in a spatial vector (density) $B^i$ and a scalar
(density) field $\phi$. The problem with these variables is that the Lagrange multipliers $N$ and $N^i$ appear explicitly
inside the constraints. This fact complicates the decoupling of constraints into first and second class.
That is why, in the second subsection, we will perform a canonical transformation to a better set of variables,
which will be called ``adapted'', for which the lapse and the shift will disappear from our constraints.
This will simplify the decoupling between first and second class constraints. In the third subsection,
the Poisson brackets between different constraints will be presented and it will be shown how, after introducing
the Dirac brackets to deal with the second-class constraints, the usual constraint algebra of general relativity is recovered.
Finally, in the last subsection, a comparison between this system and the weakly-coupled scalar-field model 
will be performed.

\subsection{Dual decomposition of the three-form field}

The three-form $A_{\mu\nu\rho}$ contains four independent components. Therefore,
it is very illustrative to perform a change of variables and encode the nondynamical part $A_{0ij}$
on a pseudo-vector (vector-density of weight $+1$) field $B^i$,
\begin{equation}
 A_{0ij}=:\eta_{ijk}B^k,
\end{equation}
with conjugate momentum $\widetilde{P}^{\vec{B}}_k$, defined by $\Pi^{0ij}=\frac{1}{6}\eta^{ijk} \widetilde{P}^{\vec{B}}_k$.
Here we have defined $\eta_{ijk}$ and $\eta^{ijk}$ as the completely antisymmetric
spatial densities (of weight $-1$ and $+1$, respectively), whose components in the chosen basis are $+1$, $-1$ or $0$. In this way,
the three primary constraints \eqref{primary1}, would read
\begin{equation}
 \widetilde{P}^{\vec{B}}_i=0.
\end{equation}
In a similar way, one can choose the pseudo-scalar (scalar-density of weight $+1$)
$\phi$ to describe the physical degree of freedom of the three-form field,
\begin{equation}\label{defphi}
 A_{ijk}=:\phi\eta_{ijk},
\end{equation}
with canonical momentum $\widetilde{\Pi}$,
\begin{equation}
\Pi^{ijk}=\frac{1}{6}\widetilde{\Pi}\,\eta^{ijk}.
\end{equation}

At this stage, one could wonder why are we choosing densities, instead of proper tensorial quantities,
to encode our variables. For instance, it might be more natural to encode the only physical
degree of freedom of $A_{ijk}$, in a scalar field $\chi$:
\begin{equation}\label{defchi}
 A_{ijk}=:\chi\epsilon_{ijk},
\end{equation}
and the conjugate momentum of such a field $\Pi_\chi$, would be defined as follows:
\begin{equation}
\Pi^{ijk}=:\frac{1}{6}\Pi_{\chi}\,\epsilon^{ijk},
\end{equation}
with $\epsilon_{ijk}=\sqrt{h}\eta_{ijk}$ and $\epsilon^{ijk}=\eta^{ijk}/\sqrt{h}$.
The problem with this choice of variables is that, since there is an implicit determinant of the metric
in the definition \eqref{defchi}, one gets
\begin{equation}
\Pi^{ijk}\partial_0A_{ijk}=\Pi_\chi\partial_0\chi+\frac{1}{2}\chi\Pi_\chi h^{ij}\partial_0h_{ij},
\end{equation}
which implies a change of the momentum of the metric $\pi^{ij}$ in order to keep the symplectic structure unchanged. 
Furthermore, in order to compare three-form models with the system given by a scalar field weakly coupled to general relativity
(as will be explicitly explored in Secs. \ref{constalg} and \ref{sec_quadratic}), it is necessary to treat the conjugate momentum
$\widetilde{\Pi}$ as the one playing the role of the scalar field. Note that, in our case, the variables $\phi$ and $B^i$
introduced above are pseudo-tensors, whereas their conjugate momenta $\widetilde{\Pi}$ and $\widetilde{P}^{\vec{B}}_i$ are proper tensors.

Therefore, in order to keep the geometric variables unchanged and to ease the comparison with the weakly-coupled scalar-field model,
we will proceed with the variables $\phi$ and $B^i$ and their conjugate momenta $\widetilde{\Pi}$ and $\widetilde{P}^{\vec{B}}_i$. 
In terms of them, the primary Hamiltonian reads as follows,
\begin{equation}
{\cal H}_p=N \left[C +\frac{\sqrt{h}}{2}\widetilde{\Pi}^2+\sqrt{h}V(A^2)\right]+N^i C_i
+\widetilde{\Pi} D_i B^i+\alpha\, \widetilde{P}^N+\alpha^i \widetilde{P}^{\vec N}_i+\beta^i \widetilde{P}^{\vec{B}}_i,
\end{equation}
where we have introduced $\beta^i:=\partial_0B^i$.
The argument of the potential is given as,
\begin{equation}\label{argument}
 A^2=\frac{6}{h N^2}\left[(N^2-N_iN^i)\phi^2+2\phi N_i B^i- B_iB^i\right],
\end{equation}
where spatial indices are contracted with the spatial metric $h_{ij}$.
Finally, the seven nontrivial constraints of the system take the following form:
\begin{eqnarray}\label{constraint1}
C+\frac{\sqrt{h}}{2}\widetilde{\Pi}^2+\sqrt{h}V+N\sqrt{h}\frac{\partial V}{\partial N}=0,\\\label{constraint2}
 C_i+N\sqrt{h}\frac{\partial V}{\partial N^i}=0,\\\label{constraint3}
 D_i\widetilde{\Pi}-N\sqrt{h}\frac{\partial V}{\partial B^i}=0.
\end{eqnarray}

\subsection{Canonical transformation to adapted variables}

As already commented above, the problem with the variables introduced in the previous section,
is that the lapse $N$ and the shift $N^i$ appear inside the constraints.
We will show that it is possible to perform a canonical transformation that removes this dependence completely.

More concretely, the potential $V=V(h_{ij},\phi, B^i,N,N^i)$ depends on the Lagrange multipliers, as can be explicitly
seen on the form of its argument \eqref{argument}. In this way, the different derivatives of the potential
in the last term of constraints \eqref{constraint1}-\eqref{constraint3},
implies the appearance of the lapse and the shift inside the constraints.
Let us write these derivatives explicitly,
\begin{eqnarray}
N \frac{\partial V}{\partial N}&=&\frac{ 12}{h N^2} (\phi^2 N^i N_i+B^i B_i-2\phi N^i B_i) V',\\
 N \frac{\partial V}{\partial N^i}&=&-\frac{12\phi}{hN}(\phi N_i-B_i) V',\\
 N \frac{\partial V}{\partial B^i}&=&\frac{12}{hN}(\phi N_i-B_i) V',
\end{eqnarray}
where $V'$ stands for the derivative of the potential with respect to its argument.
This form of the derivatives motivates the following change
of variable $B^i\rightarrow \omega^i$ defined as,
\begin{equation}
\omega^i=\frac{1}{N}(\phi N^i-B^i)\,,\qquad B^i=\phi N^i-N\omega^i\,,
\end{equation}
which will in fact absorb all the dependence of the constraints in the lapse and the shift.
In particular, now the argument of the potential takes the form
$A^2=\frac{6}{h}(\phi^2-\omega^2)$, with $\omega^2:=\omega^i\omega_i$, and thus
the potential $V=V(h,\phi,\omega)$ turns out to be a function of the determinant of the metric $h_{ij}$, the field $\phi$ and the module
of the vector $\omega^i$.
In order to preserve the symplectic structure of our theory,
this change implies also a change of some of the momenta of our
variables. More precisely, the new momenta
$(P^{\vec{\omega}}_i,P^{N}, P^{\vec{N}}_{i},\Pi)$
read as follows,
\begin{align}
P^{\vec{\omega}}_i&=-N\widetilde{P}^{\vec{B}}_i\,, & 
\widetilde{P}^{\vec{B}}_i&=-\frac{1}{N}P^{\vec{\omega}}_i\,,\\
P^{N}&=\widetilde{P}^{N}-\frac{1}{N}(\phi N^i-B^i) \widetilde{P}^{\vec{B}}_i\,, & 
\widetilde{P}^{N}&=P^{N}-\frac{\omega^i}{N}P^{\vec{\omega}}_i\,,\\
P^{\vec{N}}_i&=\widetilde{P}^{\vec{N}}_i+\phi \widetilde{P}^{\vec{B}}_i\,, &
\widetilde{P}^{\vec{N}}_i&=P^{\vec{N}}_i+\frac{\phi}{N} P^{\vec{\omega}}_i\,,\\
\Pi&=\widetilde{\Pi}+N^i\widetilde{P}^{\vec{B}}_i\,, & 
\widetilde{\Pi}&=\Pi+\frac{N^i}{N} P^{\vec{\omega}}_i\,.
\end{align}

In terms of these new variables, our constraints take the following form:
\begin{align}
& C_1:=P^N=0\,,& & C_4:=C+\frac{\sqrt{h}}{2}\Pi^2+\sqrt{h}V+\omega^iD_i\Pi=0\,,\\
& C_{2i}:=P^{\vec{N}}_i=0\,,& & C_{5i}:=C_i-\phi D_i\Pi=0\,,\\
& C_{3i}:=P^{\vec{\omega}}_i=0\,,& & C_{6i}:=D_i\Pi-\frac{12}{\sqrt{h}}V'\omega_i=0\,,
\end{align}
which do not contain the Lagrange multipliers $N$ and $N^i$.
Hence, we have seven primary constraints $(C_1,C_{2i},C_{3i})$, seven secondary constraints $(C_4,C_{5i},C_{6i})$,
and the primary Hamiltonian is written as a linear combination of them,
\begin{equation}
\mathcal{H}_p=NC_4+N^iC_{5i}+\alpha\, C_1+\alpha^i C_{2i}+\gamma^i C_{3i}\,,
\label{Ham2}
\end{equation}
where the coefficient $\gamma^i$ is defined as the time derivative of the vector field $\omega^i$ $(\gamma^i:=\partial_0\omega^i)$.

Making use of the Poisson brackets between the constraints,
that are explicitly given in the next section \eqref{bracket55}--\eqref{bracket66}, one can easily show
that the evolution of the Hamiltonian constraint $C_4$ and diffeomorphism constraints $C_{5i}$ is conserved on shell,
since the brackets $\{C_4,\mathcal{H}_p\}$ and $\{C_{5i},\mathcal{H}_p\}$ are just a linear combination of constraints.
On the contrary, the matter constraint $C_{6i}$ does not commute on-shell with the Hamiltonian,
and thus its evolution imposes conditions on the coefficient $\gamma^i$,
\begin{equation}
\{C_{6i},{\cal H}_p\}= \{C_{6i},NC_4+N^jC_{5j} \}+\gamma^j\{C_{6i},C_{3j} \}=0.
\label{fixbeta}
\end{equation}
Therefore, $\omega^i$ is not a free Lagrange multiplier, as the lapse or the shift, since
its time derivative is restricted by this last relation. As will be commented below,
this is a signal that indicates that $C_{6i}$ is not a first-class constraint, since
one is not free to choose a (gauge-fixing) condition to be fulfilled by its corresponding
Lagrange multiplier $\omega^i$.

For completeness, we provide the equations of motion for physical quantities in terms of
these adapted variables.
The evolution of $\phi$ and $\Pi$ is given by,
\begin{eqnarray}\label{evolphi}
\partial_0\phi &=& N\sqrt{h}\,\Pi+D_i\left(\phi N^i-N\omega^i\right),\\\label{evolPi}
\partial_0\Pi&=& -\frac{12 N}{\sqrt{h}} \phi V'+N^iD_i\Pi\,,
\end{eqnarray}
whereas the evolution of $h_{ij}$ and $\pi^{ij}$ reads as follows,
\begin{eqnarray}
 \partial_0 h_{ij} &=& \frac{N}{\sqrt{h}}(2\pi_{ij}-\pi h_{ij})+2 D_{(i}N_{j)},\label{evolh}\\
  \partial_0\pi^{ij}&=& -\frac{N}{2}\sqrt{h}\left(2R^{(3)ij}-R^{(3)}h^{ij}\right)
  +\frac{N}{4\sqrt{h}}h^{ij}\left(2\pi_{mn}\pi^{mn}-\pi^2\right)
  -\frac{N}{\sqrt{h}}\left(2\pi^{im}\pi_m^{j}-\pi\pi^{ij}\right)\nonumber\\ 
&+&\sqrt{h}\left(D^iD^jN-h^{ij}D^mD_mN\right)+
D_m\left(N^m\pi^{ij}\right)-2\pi^{m(i}D_mN^{j)}\nonumber\\
 &-&\frac{\sqrt{h}}{4}Nh^{ij} \left(\Pi^2+ 2 V\right)-\frac{6 N V'}{\sqrt{h}}\left[ (\omega^2-\phi^2)h^{ij}-\omega^i\omega^j \right].\label{evolpi}
\end{eqnarray}
Note that, in the last two equations, the dependence on the matter field appears only in the last line.

\subsection{Constraint algebra}\label{constalg}

We define the Poisson brackets between any two functionals on the phase space $F$ and $G$ as,
\begin{equation}
 \{F,G\}=\int d^3x\sum_I \left[\frac{\delta F}{\delta q_I}\frac{\delta G}{\delta p^I}-\frac{\delta G}{\delta q_I}\frac{\delta F}{\delta p^I}\right],
\end{equation}
where the index $I$ runs over variables $q_I=(h_{ij},\phi,\omega^i,N,N^i)$ and conjugate momenta $p^I=(\pi^{ij},\Pi,P^{\vec{\omega}}_i, P^N, P^{\vec N}_i)$.
In particular, as is well known \cite{Romano:1991up}, the constraints of vacuum general relativity
are closed under Poisson brackets:
\begin{eqnarray}\label{vacuumalgebra1}
&& \{C_i [f^i], C_j [g^j]\}= C_j[f^iD_ig^j-g^iD_if^j]\,,
\\\label{vacuumalgebra2}
&&\{C_i [f^i],C[g]\}=C[f^iD_ig]\,,
\\\label{vacuumalgebra3}
&& \{C[f],C[g]\}=C_i[fD^ig-gD^if]\,,
\end{eqnarray}
where we have defined the functional associated to any function $F$ as,
\begin{equation}
F[f]:=\int d^3x F f,
\end{equation}
with $f$ an arbitrary smearing function.

Let us now analyze how the algebra of  general relativity is modified due to the coupling of a three-form field.
On the one hand, with the adapted variables introduced in the last subsection,
it is clear that the constraints $C_1$ and $C_{2i}$ commute off-shell with all the others
because of the absence of the lapse and the shift in the constraints. Thus, these four constraints
are automatically first class.

On the other hand, the brackets between the diffeomorphism and Hamiltonian constraints, which generalize the above
expressions \eqref{vacuumalgebra1}--\eqref{vacuumalgebra3}, take the following form,
\begin{eqnarray}\label{bracket55}
\{C_{5i}[f^i],C_{5j}[g^j]\} &=& C_{5i}\left[f^jD_jg^i-g^jD_jf^i\right]\,,\\\label{bracket54}
  \{C_{5i}[f^i],C_{4}[g]\}&=&C_4[f^iD_ig]-C_{6i}[g(\omega^jD_jf^i-D_j(f^j\omega^i))]\,,\\\label{bracket44}
\{C_{4}[f],C_{4}[g]\}&=&C_{5i}[fD^ig-gD^if] + C_{6i}[\phi(fD^ig-gD^if)]\,.
\end{eqnarray}
Note that, all these brackets vanish on-shell. The difference with the vacuum algebra
\eqref{vacuumalgebra1}--\eqref{vacuumalgebra3} is the appearance of the constraint $C_{6i}$
in the brackets that involve the Hamiltonian constraint.

Regarding the primary matter constraint $C_{3i}$, it trivially commutes with itself and (weakly) with the geometric constraints,
namely,
\begin{eqnarray}\label{bracket3}
&& \{C_{3i}[f^i],C_{3j}[g^j]\}=0,\qquad\{C_{3i}[f^i],C_{5j}[g^j]\}=0\,,\qquad
\{C_{3i}[f^i],C_4[g]\}=-C_{6i}[f^i g]\,,
\end{eqnarray}
but not in the general case ($V'\neq 0$) with the other matter constraint $C_{6i}$,
\begin{equation}\label{brackets36}
 \{C_{3i}[f^i],C_{6j}[g^j]\}=\int d^3x \frac{12}{\sqrt{h}}\left(V'h_{ij}-\frac{12}{h}V''\omega_i \omega_j\right)f^ig^j\,.
\end{equation}
In fact, the matter constraint $C_{6i}$ does not commute with any constraint,
except for $C_1$ and $C_{2i}$,
\begin{eqnarray}
 \{C_{4}[f],C_{6i}[g^i]\}&=&\int d^3x\,\bigg[\frac{6}{h}fV'
\left(
4 g^i\omega^j\pi_{ij}-\pi g^i\omega_i-2\sqrt{h}\phi D_ig^i
\right)
-\frac{144}{h^{3/2}}g^i\omega_i\phi V''\omega^jD_jf
\nonumber\\\label{bracket46}
& + & \frac{72}{h^2}g^i\omega_ifV''\left(
\pi\phi^2-2\pi_{ij}\omega^i\omega^j+2h\phi\Pi-2\sqrt{h}\phi D_i\omega^i
\right)\bigg]
\,,\\
\{C_{5i}[f^i],C_{6j}[g^j]\}&=&\int d^3x \bigg[\frac{24}{\sqrt{h}}V'g^i\omega^j D_{(i}f_{j)}-\frac{12}{\sqrt{h}}V'g^i\omega_i D_jf^j
+ \frac{144}{h^{3/2}}g^i\omega_i V''
\left(
\phi f^jD_ j\phi+\omega^2D_jf^j-\omega^i\omega^jD_if_j
\right)
\nonumber\\\label{bracket56} & +& f^iD_ i\Pi D_jg^j\bigg]
\,,\\\label{bracket66}
\{C_{6i}[f^i],C_{6j}[g^j]\}&=&\int d^3x\frac{144}{h^{3/2}}\phi V''\omega_i\left(
f^iD_ jg^j-g^iD_jf^j
\right)\,.
\end{eqnarray}

In summary, $C_1$ and $C_{2i}$ are the only first-class constraints 
of this system since they strongly commute with all the other constraints.
In particular, for the general case $V'\neq 0$ (the singular
case $V'=0$ will be explored in Sec. \ref{constpot}), the Hamiltonian and diffeomorphism constraints, $C_4$ and $C_{5i}$, are
not first class due to their nonvanishing brackets with $C_{6i}$ \eqref{bracket46}--\eqref{bracket56}.

At this point one might try to construct linear combinations of constraints
in order to decouple them between first and second class. Nonetheless, in this case
this might not be possible due to several technical difficulties.
In particular, the presence of the momentum of the metric $\pi^{ij}$
(in a different form that it appears inside $C_4$) in the brackets between the constraints
$C_{6i}$ and $C_4$ \eqref{bracket46} is a crucial problem
since this kind of terms do not appear anywhere else. Therefore, it is not possible
to construct a first-class constraint by a linear combination of $C_4$ and other constraints
because when computing the Poisson brackets between such linear combination and $C_{6i}$,
one will not get the required terms to cancel out the dependence on $\pi^{ij}$.

Alternatively, one can identify the purely second-class constraints to construct a corresponding Dirac bracket.
Note that the pair $(\omega^i, P^{\vec{\omega}}_i)$ is nonphysical: the vector $\omega^i$ plays
the role of a Lagrange multiplier, but its time derivative is constrained by relation \eqref{fixbeta},
and $P^{\vec{\omega}}_i$ is a pure-constrained
variable. One can then solve some of the constraints for this unphysical couple.
The momentum is trivially obtained by solving $C_{3i}$, whereas the only constraint that allows for a resolution
for $\omega^i$ is $C_{6i}$. In addition, the Poisson brackets between these constraints, $C_{3i}$ and $C_{6i}$,
is nonvanishing, which allows us to define an invertible matrix $A_{IJ}=\{{\cal C}_I,{\cal C}_J\}$, with ${\cal C}_I$
being the different elements of $C_{3i}$ and $C_{6i}$, to construct the corresponding Dirac bracket.
In practice one just computes Poisson brackets between different objects, and then imposes constraints
$P^{\vec{\omega}}_i=0$ and the form of $\omega^i$ in terms of the rest of the variables obtained from $C_{6i}=0$.
In terms of these Dirac brackets $\{\cdot,\cdot\}_D$, one then recovers the usual algebraic structure of vacuum general relativity,
\begin{eqnarray}
\{C_{5i}[f^i],C_{5j}[g^j]\}_D &=& C_{5i}\left[f^jD_jg^i-g^jD_jf^i\right]\,,\\
 \{C_{5i}[f^i],C_{4}[g]\}_D &=&C_4[f^iD_ig]\,,\\
 \{C_{4}[f],C_{4}[g]\}_D &=& C_{5i}[fD^ig-gD^if]\,.
\end{eqnarray}
Nevertheless, the procedure to use the Dirac brackets involves solving $C_{6i}=0$ for $\omega^i$. Since $\omega^2$
appears inside the potential $V$, it is not possible to do it explicitly for generic potentials.
That is why some particular cases will be analyzed in the next section.

In summary, in four dimensions a three-form matter field $A_{\mu\nu\rho}$ contains four degrees of freedom,
which have been encoded in the pseudo-scalar field $\phi$ and the pseudo-vector field $\omega^i$, with its corresponding
conjugate momenta, the scalar field $\Pi$ and the vector field $P^{\vec{\omega}}_i$. It has been shown that, between these degrees of freedom
there is only a physical one, represented by the pair $(\phi,\Pi)$, whereas the other three pairs
$(\omega^i,P^{\vec{\omega}}_i)$ are obtained by solving the matter constraints $C_{3i}=0$ and $C_{6i}=0$.
Therefore, this theory is equivalent to a scalar field with a nontrivial coupling to general relativity.

\subsection{Comparison with weakly-coupled scalar-field model}

For comparison purposes, let us display the Hamiltonian corresponding to a scalar field $\varphi$ weakly coupled to general relativity
with a potential ${\cal V}(\varphi)$:
\begin{equation}\label{scalarfieldhamiltonian}
 {\cal H}_\varphi=N\left(C+ \frac{p_\varphi^2}{2\sqrt{h}}
                  +  \frac{\sqrt{h}}{2} \varphi_{,i}\varphi^{,i}+\sqrt{h}{\cal V}(\varphi)\right)+ N^i\left(C_i+ p_\varphi\, \varphi_{,i}\right)
                  +\alpha P^{N}+\alpha^i P^{\vec{N}}_i,
\end{equation}
with the usual convention $\varphi_{,i}:=\partial_i\varphi$.

In the case of the three-form field, due to the decomposition that has been performed, the momentum $\Pi$ plays the role
of the scalar field $\varphi$, whereas the variable $\phi$ plays the role of the conjugate momentum $p_\varphi$.
Therefore, in order to compare the Hamiltonian of the three-form field \eqref{Ham2} with that of the weakly-coupled scalar field \eqref{scalarfieldhamiltonian},
one should perform the following changes: $\Pi\rightarrow-\varphi$ and $\phi\rightarrow p_\varphi$. By doing this,
the Hamiltonian \eqref{Ham2} takes the following form: 
\begin{equation}\label{3formfieldhamiltonian}
 {\cal H}_p=N\left(C+\sqrt{h}\, V(h,p_\varphi,\omega)-\omega^i \varphi_{,i}+ \frac{\sqrt{h}}{2}\varphi^2\right)
 + N^i\left(C_i+ p_\varphi\, \varphi_{,i}\right)+\alpha P^{N}+\alpha^i P^{\vec{N}}_i,
\end{equation}
where, for convenience, $C_{3i}=0$ has already been imposed and the vector $\omega^i$ should be obtained from
$C_{6i}=0$ and thus $\omega^i=\omega^i(p_\varphi\,,p_{\varphi,j}\,,h)$, which introduces a dependence on the gradient
of the momentum of the scalar field in the potential $V$.
Note that the diffeomorphism constraint is exactly the same for both models, but not the Hamiltonian constraint in the general case.
In fact, by direct comparison between \eqref{scalarfieldhamiltonian}
and \eqref{3formfieldhamiltonian}, one can qualitatively state that a three-form model is equivalent to a weakly-coupled
scalar-field model with quadratic potential ${\cal V}$. Nonetheless, in this case the precise form of the kinetic term
would depend on the explicit form of the potential $V$. In particular, for the case that the potential of the three-form $V$ is also quadratic,
one could exactly recover the Hamiltonian of a weakly-coupled scalar field from the three-form model.
Even so, there are some slight differences between the equations of motion, as will be analyzed in detail in Sec. \ref{sec_quadratic} below.
Therefore, even if a three-form matter field is equivalent to a scalar field in the sense that it contains
one degree of freedom, in the general case its dynamics is quite different from a weakly-coupled scalar field.

\section{Application to particular cases}\label{cases}

In the previous sections, a completely generic model of a three-form matter field coupled to general relativity has been considered.
Up to now, neither specific form of the potential $V$ nor spacetime symmetries have been imposed.
Nevertheless, as it was stated in the previous section, for general models, the explicit dependence of the potential on its argument $A^2$
is necessary to further develop the equations.

In this section different particular cases will be analyzed.
More concretely, after studying the case of the constant potential, in Subsec. \ref{sec_quadratic} we will explore
the interesting case of the quadratic potential. In fact, it will be shown that this case is quite similar to a weakly-coupled scalar-field
model, also with a quadratic potential. Finally, in Subsec. \ref{cosmology}, we will explicitly comment on the cosmological case
of an homogeneous and isotropic universe widely treated in the literature.

\subsection{Constant potential}\label{constpot}

The singular case of a constant potential $V=V_0$ is not included in the general discussion presented above.
As a consequence of the fact that, in this case, the derivative of the potential is vanishing, the matter constraint $C_{6i}$ simplifies to $C_{6i}=D_i\Pi=0$.
In addition, and more importantly, now it strongly commutes with all the constraints \eqref{brackets36}--\eqref{bracket66}. In fact, not only $C_{6i}$, but 
all the constraints turn out to be first class. Furthermore, the evolution of $C_{6i}$, as can be seen in \eqref{evolPi}, requires that the momentum $\Pi$ be time independent.

In order to understand better the dynamics of this system, it is convenient to impose the following gauge-fixing condition for $C_{6i}$,
\begin{equation}\label{gaugefixing}
 \phi=0,
\end{equation}
which does not commute with $C_{6i}$. Now this constraint is second-class and, solving it, one straightforwardly gets that $\Pi$ is an arbitrary constant $\Pi=k$
fixed by initial conditions. In addition, from equation \eqref{evolphi}, the evolution of this gauge-fixing condition implies the following equation
\begin{equation}
 N\sqrt{h}k-D_i(N\omega^i)=0,
\end{equation}
to be solved for the vector $\omega^i$. Finally, in order to complete the gauge-fixing procedure, one also strongly imposes $P^{\vec{\omega}}_i=0$ everywhere.

In this way, one ends up with the same degrees of freedom as vacuum general relativity, characterized by the dynamical couple $(h_{ij},\pi^{ij})$
and the Lagrange multipliers $N$ and $N^i$. The only terms of the Hamiltonian \eqref{Ham2} that contribute to the evolution of these variables
read as follows,
\begin{equation}
\mathcal{H} = N (C+\sqrt{h}\Lambda)+N^iC_i\,,
\end{equation}
where $\Lambda:=V_0+k^2/2$ plays the role of a cosmological constant. Concerning the evolution equations, note that
this constant only enters in the evolution equation for the momentum of the metric $\pi^{ij}$ given by \eqref{evolpi}.

In summary, the case of a three-form field with a constant potential resembles exactly vacuum general relativity with a cosmological constant.

\subsection{Quadratic potential}\label{sec_quadratic}

We will consider now the special case when
the potential is linear in its argument $A^2$ and, thus, it is
quadratic in the variables $\phi$ and $\omega$. Its explicit form is then given by
\begin{equation}
 V=A^2=\frac{6}{h}(\phi^2-\omega^2).
\end{equation}
In this case, one can explicitly follow the general procedure explained in Sec. \ref{constalg}
of solving the constraints $C_{3i}=0$ and $C_{6i}=0$.
Hence, one straightforwardly obtains the following expressions for the couple ($\omega^i$, $P^{\vec{\omega}}_i$):
\begin{equation}\label{replacement}
\omega_i=\frac{\sqrt{h}}{12 }D_i\Pi\,,\qquad P^{\vec{\omega}}_i=0,
\end{equation}
which should be imposed in the rest of the constraint and evolution equations.

With the goal of exploring the analogies between this model and the system of a scalar field weakly coupled to gravity,
we will impose the replacements \eqref{replacement}, and perform the change of variables
$\Pi\rightarrow-2\sqrt{3}\varphi$ and $\phi\rightarrow p_\varphi/(2\sqrt{3})$ in the Hamiltonian \eqref{Ham2}.
In this way, one gets the following Hamiltonian,
\begin{equation}
 {\cal H}_p=N\left(C+ \frac{p_\varphi^2}{2\sqrt{h}}
                  +  \frac{\sqrt{h}}{2} \varphi_{,i}\varphi^{,i}+6\sqrt{h}\varphi^2\right)+ N^i\left(C_i+ p_\varphi\, \varphi_{,i}\right)
                  +\alpha P^{N}+\alpha^i P^{\vec{N}}_i.
\end{equation}
Note that this Hamiltonian can not be used as the generator of the evolution of the system, since two constraints have
already been solved \eqref{replacement}.
In particular, this Hamiltonian is exactly the same as the one of the weakly-coupled scalar field \eqref{scalarfieldhamiltonian},
given that one chooses the potential ${\cal V}=6\varphi^2$ there. Even if the Hamiltonian is formally the same as for
the scalar field, the equations of motion might (and, in fact, do) differ since the computation of Poisson
brackets do not commute with the resolution of the second-class constraints.

In fact, the difference between the two models is quite slight. As commented, the constraint equations are exactly the same.
Regarding the evolution equations, the equations for
the matter part $(\varphi, p_\varphi)$ also turn out to be the same. In addition, the equation of motion for
the three-metric \eqref{evolh} does not contain any contribution from matter variables, so it is also exactly
equal for both models. Finally, the evolution equation for the conjugate momentum of the metric
$\pi^{ij}$ \eqref{evolpi} is the only one that presents a difference between the two models.
Performing the commented changes, the last line of equation \eqref{evolpi} reads as,
\begin{equation}
-\frac{N}{4 \sqrt{h}} h^{ij}\left(12 h \varphi^2 -p_\varphi^2+h\varphi_{,k}\varphi^{,k}\right) +\frac{N}{2}\sqrt{h}\, \varphi^{,i} \varphi^{,j} .
\end{equation}
While in the case of a weakly-coupled scalar field with potential ${\cal V}=6\varphi^2$
these terms would be as follows:
\begin{equation}
-\frac{N}{4 \sqrt{h}} h^{ij}\left(12 h \varphi^2 -p_\varphi^2+h\,\varphi_{,k}\varphi^{,k}\right) .
\end{equation}
Therefore, the only difference between the two models is the presence of the term $\frac{N}{2}\sqrt{h}\, \varphi^{,i} \varphi^{,j}$ in
the equation of motion of $\pi^{ij}$ for the case of a three-form matter field, that does not appear for
the weakly-coupled scalar field. Whenever this term can be neglected (as can be exactly done, for instance, in homogeneous models)
the dynamics of both systems will be the same.

\subsection{Homogeneous and isotropic models}\label{cosmology}

As it was commented in the introduction, one of the main applications of the three-form models during the last few years has been developed within a cosmological framework.
In a spacetime given by the Friedmann-Lemaitre-Robertson-Walker metric the spacial vector $\omega_i$, as well as all spatial derivatives, must be vanishing
due to homogeneity and isotropy.
This fact simplifies enormously the set of constraints presented above, and most of them are trivially obeyed.
The only nontrivial constraint is the Hamiltonian constraint $C_4$.

Assuming a spatially flat model, $R^{(3)}=0$, with metric
\begin{equation}
ds^2=-dt^2+a(t)^2 d\vec{x}^2\,,
\end{equation}
the Hamiltonian constraint reads as follows,
\begin{equation}
 C_4=\frac{1}{\sqrt{h}}\pi^{ij}\pi_{ij}-\frac{\pi^2}{2\sqrt{h}}+\frac{\sqrt{h}}{2}\Pi^2+\sqrt{h} V=0\,.
\end{equation}
In these models the natural variables are the scale factor $a$ and its canonical conjugate
momentum $p_a$. Therefore, taking into account that $h=a^6$ and $h_{ij}=a^2\delta_{ij}$,
one writes $\pi^{ij}=\frac{p_{a}}{2a}\delta^{ij}$. In this way, the above constraint takes the following form,
\begin{equation}
C_4=-\frac{1}{24a}p_a^2+\frac{a^3}{2}\Pi^2+a^3 V=0.
\end{equation}
Hence, we recover the Hamiltonian used for the cosmological three-form model  \cite{BouBriGar}.

\section{Conclusions}\label{conclusions}

In this paper we have analyzed a three-form matter field coupled with general relativity in four dimensions.
A Legendre transformation has been performed to obtain the corresponding Hamiltonian of the system. As usually happens
in gauge theories, this step has defined some primary constraints. By evolution of them, the complete set of constraints of this model has been
obtained. They are quite entangled, not even one of them being first class.
In part, this is due to the fact that the Lagrange multipliers appear inside the constraints.
Therefore, a canonical transformation has been proposed, which makes all of them independent of
the lapse and the shift, in order to decouple some of them as first class. At this point,
the algebra of the constraints has been analyzed, by explicitly computing the Poisson brackets
between any pair of them. The results show that there are some second-class constraints,
which have been dealt with by considering the Dirac bracket.

In this way, the duality
between a three-form matter model and a scalar field has been explicitly shown. These two models contain the same
physical degrees of freedom, since in four dimensions a three-form field $A_{\mu\nu\rho}$
contains four independent components, which have been encoded in the scalar $\phi$
and the vector $\omega^i$, and their corresponding conjugate momenta, $\Pi$ and $P^{\vec{\omega}}_i$ respectively.
However, the vector $\omega^i$ turns out not to contain any physical information, since
its conjugate momentum is constrained to be vanishing, and the vector itself must be explicitly obtained
in terms of other variables when computing the Dirac brackets. 
In summary, all the physical information of the three-form field is encoded in the couple $(\phi,\Pi)$.

In fact, looking at their corresponding Hamiltonians, \eqref{scalarfieldhamiltonian} and \eqref{3formfieldhamiltonian},
one could say that a three-form model is equivalent to a weakly-coupled
scalar-field model with quadratic potential ${\cal V}$, although with a different kinetic term
that depends on the explicit form of the potential $V$. This fact, obviously, might produce
crucial differences between the evolutions of both systems.

Interestingly, as explicitly shown in Sec. \ref{sec_quadratic},
the particular case of a three-form field with a quadratic potential is almost identical
to a weakly-coupled scalar field also with a quadratic potential. In fact, the Hamiltonian
and, thus, all constraints have exactly the same form, as well as the evolution equations
for the matter degrees of freedom $(\phi,\Pi)$, and for the spatial metric $h_{ij}$.
The only difference between the two models is a quadratic term in derivatives of the scalar
field that appears in the equation of motion for the conjugate momentum of the metric $\pi^{ij}$
in the three-form matter model, which is not present in the weakly-coupled scalar-field case.
For physical scenarios where this term is negligible both systems will describe the very same
dynamics. This is the case, for instance, in homogeneous cosmological models where
this term will be exactly vanishing.

The development of the Hamiltonian formalism, with the complete study of the constraints,
presented in this paper paves the way to a number of applications. On the one hand, it might
be used for generalizing homogeneous cosmological models, with three-form matter content,
to inhomogeneous cosmological scenarios, in particular by considering perturbations of the homogeneous case.
On the other hand, it could shed light on the quantization procedure.
In fact, for the homogeneous cosmological three-form model it was already successfully performed within the Wheeler-DeWitt approach \cite{BouBriGar}.
Nevertheless, the quantization of the general case is much more cumbersome and
requires further investigation.


\acknowledgments
We thank Mariam Bouhmadi-L\'opez, Jes\'us Ib\'a\~nez and Jo\~ao Morais for fruitful discussions at early stages of the project.
The authors acknowledge financial support from
project FIS2017-85076-P (MINECO/AEI/FEDER, UE), and
Basque Government Grant No.~IT956-16. This paper is based upon work from COST action CA15117 (CANTATA), supported by COST (European Cooperation in Science and Technology).



\end{document}